%% file: main.tex
\documentclass[twocolumn,10pt]{article}
\usepackage[letterpaper,left=1.95cm,right=1.95cm,top=2.5cm,bottom=3.16cm]{geometry} 
\input{Styles_Eje.sty}  

{\title{\fontsize{16}{16}
\vspace*{-0.7cm}
\textbf{
Continuous variable quantum MNIST classifiers
} \\[0.2cm]
\fontsize{12}{12}
\textbf{Classical-quantum hybrid quantum neural networks}\\[0.2cm]}
}

\author[1]{\fontsize{10pt}{10pt}\selectfont \textbf{Sophie Choe}}

\affil[1]{
\fontsize{8pt}{8pt}\selectfont Electrical and Computer Engineering, Portland State University, Portland, OR}

\begin{document}
\input{Abstract}

\subfile{intro}

\subfile{qc}

\subfile{QNN}

\subfile{exp}

\subfile{conc}

\bibliographystyle{plain}

\input{ref}
\end{document}

%% file: Abstract.tex
\twocolumn
[
\begin{@twocolumnfalse}
\maketitle
\begin{abstract}
\vspace*{0.5cm}
\justify
\fontsize{10pt}{10pt}\selectfont
In this paper, classical and continuous variable (CV) quantum neural network hybrid multi-classifiers are presented using the MNIST dataset. The combination of cutoff dimension and probability measurement method in the CV model allows a quantum circuit to produce output vectors of size $n^m$ where $n=$cutoff dimension and $m=$the number of qumodes. They are then translated as one-hot encoded labels, padded with $n^m - 10$ zeros. The total of eight different classifiers are built using $2, 3, \ldots, 8-$ qumodes, based on the binary classifier architecture proposed in "Continuous variable quantum neural networks" \cite{cv_19}. The displacement gate and the Kerr gate in the CV model allow for the bias addition and non-linear activation components of classical neural networks of the form $L(x) = \phi(Wx+b)$ to be directly implemented in quantum as $L(\ket{x})=\ket{\phi(Wx+b)}$. The classifiers are composed of a classical feed-forward neural network, a quantum data encoding circuit, and a CV quantum neural network circuit.  On a truncated MNIST dataset of 600 samples, a 4-qumode hybrid classifier achieves $100\%$ training accuracy.

\keywordsEng{quantum computing, quantum machine learning, quantum neural networks, continuous variable quantum computing, photonic quantum computing, classical quantum hybrid model, quantum MNIST classification}
\end{abstract}
\vspace{0.5cm}

\hspace*{0.7cm}
\textit{Dated: April 2022} \\
\hspace*{0.7cm}
\vspace*{0.5cm}

\end{@twocolumnfalse}
]

%% file: intro.tex
\section{INTRODUCTION}

\justifying
Unlike the original assumption that quantum computers would replace classical computers, quantum processing units (QPUs) are emerging as task-specific special purpose processing units much like Graphics Processing Units. The currently available QPUs are called near-term quantum devices because they are not yet fully fault-tolerant and are characterized by shallow and short quantum circuits. Nonetheless, the availability of these devices allows for active research on quantum algorithms specific for these devices, especially in quantum chemistry, Gaussian boson sampling, graph optimization, and quantum machine learning.

The QPUs are based on two different theoretical models of quantum computing: the discrete variable (qubit-based) model and the continuous variable (CV) model \cite{qi_12}. The discrete variable model is an extension of the computational space from $\{0,1\}$ to a complex projective Hilbert space $\mathbb{C}P^1$, which is the surface of a projective sphere. The computational basis is composed of two elements $\{\ket{0}, \ket{1}\}$ \cite{qi_10}. The CV model is an extension of the computational space to an infinite dimensional Hilbert space whose computational basis is infinite, i.e., $\{\ket{0}, \ket{1}, \ldots, \ket{n}, \ldots\}$ \cite{qi_12}. The quantum state of an information channel (qumode) is represented by an infinite complex projective linear combination of these basis states. In practice, it is approximated by by a finite linear combination using only a finite basis $\{\ket{0}, \ket{1}, \ldots, \ket{n}$. The number of basis states used for the approximation is called cutoff dimension. On an $m-$qumode system with cutoff dimension $n$, the computational state space is of dimension $n^m$. By varying cutoff dimension and the number of qumodes used for computation, we can control the dimension of the computational space.

Implementing machine learning algorithms on quantum computers is an active areaa of research. Quantum machine learning algorithms can be implemented on variational quantum circuits with parametrized quantum gates \cite{circuit_18}. A quantum circuit is a collection of quantum gates and the change of states induced by the circuit on the initial quantum state is considered quantum computation. The results of quantum computing extracted via measurement are incorporated into optimization and parameter update computations performed on classical circuits \cite{circuit_18}. 

Quantum neural networks (QNN), a subset of quantum machine learning, follow the same architectural frame work: QNN on a QPU and optimization on a CPU. The main components of classical neural networks described as $L(x)=\phi(Wx+b)$ are the affine transformation $Wx+b$ and the non-linear activation function $\phi(\cdot)$. In the qubit model, all the available unitary gates are linear in nature hence a direct implementation of bias addition and non-linear activation function is not feasible. In the CV model, the displacement gate implements bias addition and the Kerr gate, non-linear activation function, hence a direct translation of $L(x)=Wx+b$ to its quantum version $L(\ket{x})= \ket{\phi(Wx+b)}$ is naturally embedded in the model.

The proposed CV MNIST classifiers are built on Xanadu's X8 simulator, which simulates quantum computation on an $8-$qumode photonic quantum computer. The ability to control the size of output vectors based on the number of qumodes and the notion of cutoff dimension allows for producing one-hot encoded labels of MNIST dataset. Different architectures on $2-$qumodes, $3-$qumodes, up to $8-$qumodes are introduced. They are classical-quantum hybrid networks with classical feed-forward neural networks, quantum data encoding circuits, and CV quantum neural network circuits according to the CV binary classifier proposed by Killoran et al \cite{cv_19}. The quantum machine learning software library, Pennylane \cite{Penny}, is used for the quantum circuit and Tensorflow Keras is used for the classical circuit and optimization. The classifiers achieve above $95\%$ training accuracy. 

This paper is organized in the following manner: In Section 2, the CV model of quantum computing is discussed, especially the infinite dimensionality of the computation state space and how the notion of of cutoff dimension allows us to determine the dimension of the actual computational state. In Section 3, we examine how CV quantum neural networks naturally implement classical neural networks. In Section 3.2, the architecture of the binary classifier in "Continuous variable quantum neural networks" is examined \cite{cv_19}. In Section 4, the structure of the 8 hybrid classifiers and the experimental results are presented.

%% file: qc.tex
\section{CV QUANTUM COMPUTING}
\begin{figure}[b]
    \centering
    \includegraphics[scale=0.8]{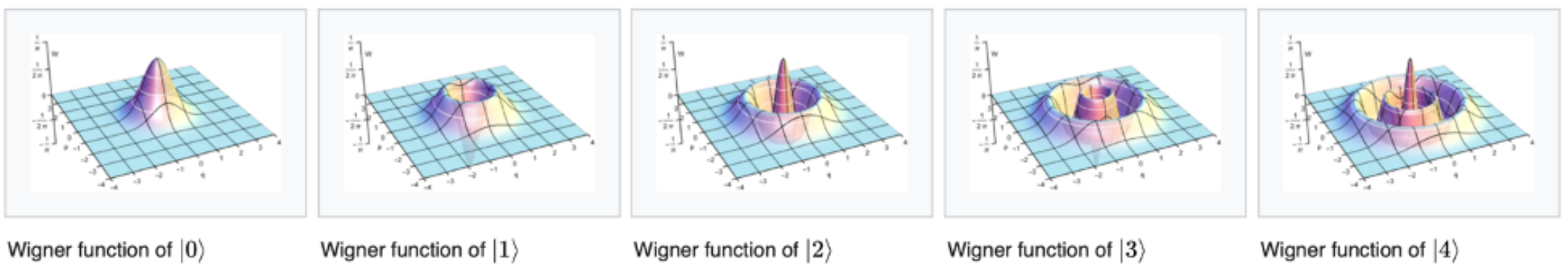}
    \caption{Fock basis: Gaussian states \href{https://upload.wikimedia.org/wikipedia/commons/7/71/Wignerfunction_fock_0.png}{image source}}
    \label{fig:Fock}
\end{figure}
The quantum state space of the CV model is an infinite dimensional Hilbert space, which is approximated by a finite dimensional space for practical information processing purposes. Under phase space representation of the model, the computational basis states are Gaussian states as opposed to single points. Therefore the quantum gates taking these Gaussian states to Gaussian states offer richer arrays of transformations than just unitary (linear) transformations as in the qubit model.

The CV model is based on the wave-like property of nature. It uses the quantum state of the electromagnetic field of bosonic modes (qumodes) as information carriers \cite{qi_cv_03}. Its physical implementation is done using linear optics, containing multiple qumodes \cite{optics_01}. Each qumode contains a certain number of photons probablistically, which can be manipulated with quantum gates.

A photon is a form of electromagnetic radiation. The position wave function $\Psi(x)$ describes the light wave electromagnetic strength of the qumode depending on the number of photons present. It is a complex valued function on real valued variables: $\Psi: \mathbb{R} \rightarrow \mathbb{C}: x \mapsto \alpha$, where $x$ is the position(s) of the photons. The wave function with more than one photon displays the constructive and destructive interactions between the light waves of the photons. 

Phase space representation of the quantum state of a linear optical system describes the state using the position $x$ and momentum $p$ variables of the photons in the given qumode system. The quasi-probablity distribution of $x$ and $p$ are plotted on the $xp-$plane, using the Wigner function. It is given by
\[
    W(x,p)=\frac{~{p}}{h}=\frac{1}{h} \int_{-\infty}^{\infty}e^{-\frac{ipy}{\hbar}}\Psi \left(x+\frac{y}{2}\right)\Psi^{\ast}\left(x-\frac{y}{2}\right) dy
\]
where $h=6.62607015 \times 10^{-34}$ is the Plank constant and $\hbar = 6.582119569 \times 10^{-16}$ the reduced Plank constant \cite{kitten_08}.

This Wigner function is applied to the position wave function $\Psi_k(x)$, where $k$ denotes the number of photons present in a qumode, to create Fock basis, also known as number basis. The image of the Wigner functions $W_0(x_0,p_0), W_1(x_1,p_1), \ldots, W_5(x_5, p_5)$ where $x_k$ and $p_k$ represent the position and momentum variables with $k$ photons present in the system is shown in Figure 1. 

\newpage
They are used as computational basis states expressing the quantum state of a system. The quantum state $\ket{\psi}$ of a qumode is expressed as a superposition of Fock basis states: 
\[
 \ket{\psi}=c_0 \ket{0}+c_1 \ket{1}+\hdots+c_n \ket{n}+\hdots, \hspace{2pt} \sum_{k=0}^{\infty} \Vert c_k \Vert ^2=1
\]

where $c_k$ is the probability amplitude of basis $\ket{k}$. 

In practice, there are not going to be an infinite number of photons physically present in a qumode. We approximate $\ket{\psi}$ with $\ket{\hat{\psi}}$ by cutting off the trailing terms. The number of Fock basis states we use to approximate the true state $\ket{\psi}$ is called "cutoff dimension". Let $n$ be cutoff dimension. Then the approximating state $\ket{\hat{\psi}}$ is in superposition of $\ket{0}, \ket{1}, \ldots, \ket{n-1}$, i.e.,
\[
 \ket{\psi}=c_0 \ket{0}+c_1 \ket{1}+\hdots+c_{n-1} \ket{n-1} \text{where} \sum_{k=0}^{n-1} \Vert c_k \Vert ^2=1.
\]
In vector representation, the state is expressed as a column vector of size $n=$cutoff dimension, with the $k^{th}$ entry being $c_k$. 

A multi-qumode system is represented by the tensor product of individual qumode states: $\ket{\psi_0} \otimes \ket{\psi_1} \otimes \ldots \otimes \ket{\psi_{m-1}}$, where $m$ is the number of qumodes. When these states are approximated by cutoff dimension $n$, the computational basis of the resulting quantum state is of size $n^m$. 
\[
\begin{aligned}
   & \ket{\psi_0} \otimes \ket{\psi_1} \otimes \ldots \otimes \ket{\psi_{m-1}}=\\ 
   &d_0 \ket{00 \ldots 0} + d_1 \ket{00 \ldots 1} + d_{n^m-1}\ket{n-1,n-1,\ldots,n-1} 
\end{aligned}
\]
Fock basis states constitute the eigenstates of the number operator $\hat{n} := \hat{a}^{\dagger}\hat{a}$ where $\hat{a}^{\dagger}$ is called the constructor and $\hat{a}$ is called the annihilator \cite{qi_12}. Their matrix representation is given by
\[
\hat{a}^{\dagger} = 
\begin{bmatrix} 
0 & 0 & 0 & \hdots & 0 & 0\\
\sqrt{1} & 0 & 0 & \ldots & 0 & 0\\
0 & \sqrt{2} & 0 & \ldots & 0 & 0\\
0 & 0 & \sqrt{3} & \ldots & 0 & 0\\
\vdots &  & \ldots & \ddots &  & \vdots\\
0 & 0 & 0 & \ldots & \sqrt{n-1}& 0\\
\end{bmatrix}
\text{and}
\]
\[
\hat{a} = 
\begin{bmatrix} 
0 & \sqrt{1} & 0 &0 & \ldots &  0\\
0 & 0 & \sqrt{2}& 0 & \ldots  & 0\\
0 & 0 & 0& \sqrt{3} & \ldots  & 0\\
\vdots &  & \ldots & \ddots &  & \vdots\\
0 & 0 & 0 & 0 &\ldots & \sqrt{n-1}\\
0 & 0 & 0 & 0 &\ldots &  0\\
\end{bmatrix}
\]

Note that the constructor $\hat{a}^{\dagger} \ket{k} = \sqrt{k+1} \ket{k+1} \text{ for }k \ge 0$ indeed constructs the next level Fock basis state and the annihilator $\hat{a} \ket{0} = 0$, $\hat{a} \ket{k} = \sqrt{k} \ket{k-1} \text{ for }k \ge 1$ annihilates the the current level Fock basis state to the lower.

The product of $\hat{a}^{\dagger}$ and  $\hat{a}$ returns a matrix 
\[
   \begin{bmatrix} 
0 & 0 & 0 &0 & \ldots &  0\\
0 & 1 & 0 & 0 & \ldots  & 0\\
0 & 0 & 2& 0 & \ldots  & 0\\
\vdots &  & \ldots & \ddots &  & \vdots\\
0 & 0 & 0 & 0 & n-2 & 0\\
0 & 0 & 0 & 0 & 0 & n-1
\end{bmatrix} 
\]
which we denote as "number operator" $\hat{n}$. The Fock basis states $\ket{k}$ are indeed eigenstates of the number operator as shown in  $\hat{n}\ket{k} = k \ket{k}$ where $k \in \{ 0, 1, \ldots, n-1 \}$ for cutoff dimension$=n$.  

Standard Gaussian gates in the CV model are of the form $U= e^{-iHt}$ where $H$ represents the Hamiltonian of the system, describing the total energy as the sum of kinetic and potential energy. In the finite quantum state space of a qumode approximated by cutoff dimension $n$, this infinite sum representing a quantum gate is also approximated by a finite matrix exponential.
The matrix exponential $e^{-iHt}$ is of the form
\[
\hat{U} = \sum_{k=0}^{n-1}\frac{({-iHt})^k}{k!}=I -iHt + \frac{(-iHt)^2}{2!}+ \hdots + \frac{{-iHt}^{n-1}}{(n-1)!}
\]

Some of the standard Gaussian gates taking Gaussian states to Gaussian states are listed below.

Squeezer with parameter $z: S(z) = exp \left( \frac{z^{\ast} \hat{a}^2+z \hat{a}^{{\dagger}^2}}{2} \right)$

Rotation with parameter $\phi: R(\phi) = exp\left(i \phi \hat{a}^{\dagger}\hat{a} \right)$

Displacement with parameter $\alpha: D(\alpha)=exp\left(\alpha \hat{a}^{\dagger}-\alpha^{\ast} \hat{a} \right)$

Another Gaussian gate, the beamsplitter, is a two-qumode gate with parameters $\theta$ and $\phi$: 
\[
B(\theta, \phi) =
exp\left( \theta \left( e^{i\phi} \hat{a} \hat{b}^{\dagger} + e^{-i \phi}
\hat{a}^{\dagger}\hat{b} \right) \right)
\]
where $\hat{b}^{\dagger}$ and $\hat{b}$ are constructor and annihilator of the 2nd qumode respectively.

Measurement is done via counting the number of photons present in each qumode with a photon detector. Xanadu's Pennylane offers a suite of measurement methods as outlined in Table 1. The size of the output vectors are affected by the number of qumodes used for computation and the notion of cutoff dimension, denoted by $n$. Currently available QPU by Xanadu, $X8$, offers $8$ qumodes. 

\begin{table}
\centering
\begin{tabular}[b]{lcc}
\hline
Measurement method & output size & output size \\
& per qumode & for $m-$qumodes \\ \hline
expectation value & $1$ & $m$\\ 
variance   & $1$ & $m$\\
probability & $n$ & $n^m$\\
\hline
\end{tabular}
\end{table}

The measurement methods used in the proposed MNIST classifiers are the expectation value method and the probability memthod.

The expectation value method produces a single real-valued output. Let $\ket{\psi} = \begin{bmatrix} \psi_0 \\ \psi_1 \end{bmatrix}$ be the quantum state of a multi-qumode system after desired quantum computational operations are performed. The expectation value measurement method returns $\bra{\psi} A \ket{\psi}$ where the operator $A$ is usually the Pauli$-X$, Pauli$-Y$, or Pauli$-Z$ gate. The expectation value of the Pauli$-X$ matrix is
\[
\begin{aligned}
    \bra{\psi} X \ket{\psi} &= 
    \begin{bmatrix} \psi_0^{\ast} & \psi_1^{\ast} \end{bmatrix}
    \begin{bmatrix} 0 & 1 \\ 1 & 0 \end{bmatrix}
    \begin{bmatrix} \psi_0 \\ \psi_1 \end{bmatrix} \\
    &= 
    \begin{bmatrix} \psi_0^{\ast} & \psi_1^{\ast} \end{bmatrix}
    \begin{bmatrix} \psi_1 \\ \psi_0 \end{bmatrix}\\
    &= 
    \psi_0^{\ast} \psi_1 + \psi_1^{\ast} \psi_0 \\
    &=
    2\left( Re(\psi_0)Re(\psi_1) + Im(\psi_0)Im(\psi_1) \right)\in \mathbb{R},
\end{aligned}
\]
which is a real number. In a multi-qumode system with $m$ qumodes, we can get a vector of size $m$ by getting the expectation value for each qumode 

The probabilities method returns the probability of each computation basis state. Suppose an $m-$qumode system has cutoff dimension $n$ for each qumode. Then, there are $n^m$ computational basis states, hence we get a vector of size $n^m$.

%% file: QNN.tex
\section{QUANTUM NEURAL NETWORKS}

Machine learning is a way of extracting hidden patterns from data by learning a set of optimal parameters for a mathematical expression that most closely match the data. The mathematical expression used for pattern extraction is called a machine learning algorithm. An algorithm with an optimal set of parameters, learned via training, is called a model. With near-term devices available on cloud, execution of quantum machine learning (QML) algorithms on quantum computers or simulators is now feasible. 

QML algorithms are built with variational circuits i.e., parametrized circuits, composed of quantum gates whose actions are defined by parameters. Training is the process of "learning" optimal parameters of the gates which produce as accurate inferences as possible for new data samples. The measurement results from a variational circuit run on a QPU are sent to a CPU for parameter optimization, i.e., computation of objective function, gradients, and new parameters. The updated parameters are fed back to the quantum circuit to adjust the parameterized gates for next iteration. The illustration of the process is shown in the figure.

\begin{figure}[H]
    \centering
    \includegraphics[scale=0.47]{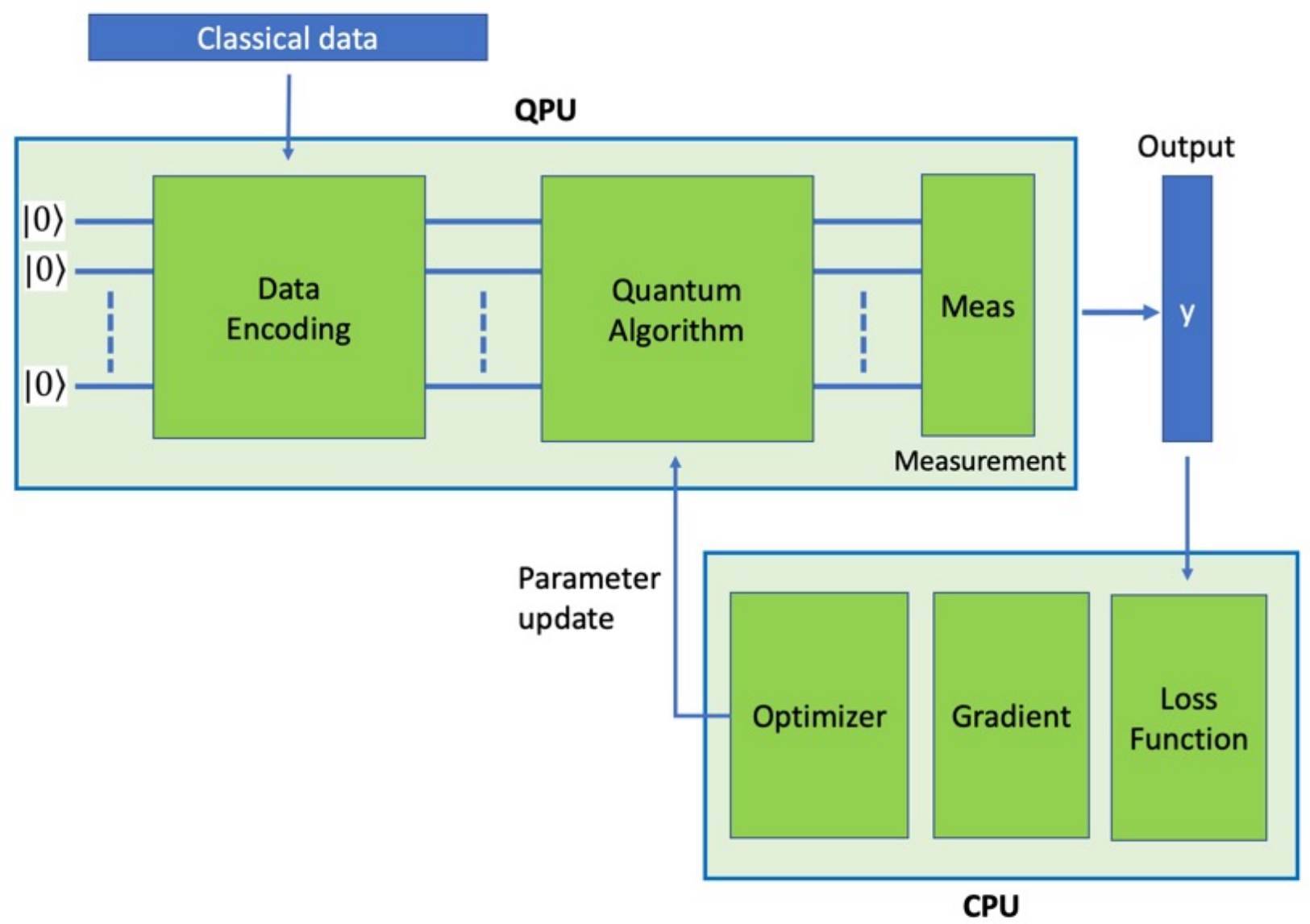}
    \caption{Variational quantum circuit parameter update}
    \label{fig:var}
\end{figure}

Google and Xanadu offer Python based software packages specifically for quantum machine learning: Tensorflow Quantum (Google) and Pennylane (Xanadu) \cite{Penny}.

Neural network is one of the subsets of machine learning algorithms, defined by a stack of layers, each composed of an affine transformation $Wx+b$ and a non-linear activation function $\phi(\cdot)$. Each layer of a neural network can be mathematically described as $L(x) = \phi(Wx+b)$. The output from one layer is then fed as input into the subsequent layer and the entire network is a composition of different layers: $L(x) = L_m \circ L_{n-1} \circ \ldots L_1(x)$. The entries of the matrix $W$ and the bias vector $b$ for each layer are learned as parameters through an iterative training process given an objective function. The goal is to find an optimal set of parameters $\{W_1, W_2, \ldots, W_m, b_1, b_2, \ldots, b_m \}$ for a network of $m$ layers. 

In quantum neural networks, the objective is to implement the classical mathematical expression $L(x)=\phi(Wx+b)$ as a quantum state $L(\ket{x})=\ket{\phi(Wx+b)}$. In converting classical neural networks into quantum circuits, the key components are:
\begin{itemize}
    \item data encoding: $x \rightarrow \ket{\psi(x)}$
    \item affine transformation $W\ket{\psi(x)}+\ket{b}$
    \item non-linear activation function $\phi(\ket{\cdot})$
\end{itemize}
In the qubit model, all available unitary gates are linear. Hence a direct way of implementing the bias addition component and the non-linear activation function component of classical neural networks into quantum is absent in the model. 

In the CV model, however, the displacement gate and the Kerr gate allow for a direct translation from classical to quantum. 

\subsection{Continuous variable QNN}

Naturally embedded in the CV model are quantum gates to directly implement the expression $L(\ket{x})=\ket{\phi(Wx+b)}$. 

The affine transformation $Wx + b$ is implemented by the composition $\mathit{D} \circ \mathit{U}_2 \circ \mathit{S} \circ \mathit{U}_1$, where $U_k$ denotes the $k^{th}$ interferometer, $S$ a set of $m$ squeezers, $D$ a set of $m$ displacement gates. The activation function $\phi(\ket{\cdot})$ is implemented by a set of Kerr gates, which are non-linear. The composition $\phi \circ \mathit{D} \circ \mathit{U}_2 \circ \mathit{S} \circ \mathit{U}_1$ acting on a quantum state $\ket{x}$ gives us the desired state $L(\ket{x})= \ket{\phi(Wx+b)}$. The schematic of the circuit is shown below.
\begin{figure}[H]
    \centering
    \includegraphics[scale=0.38]{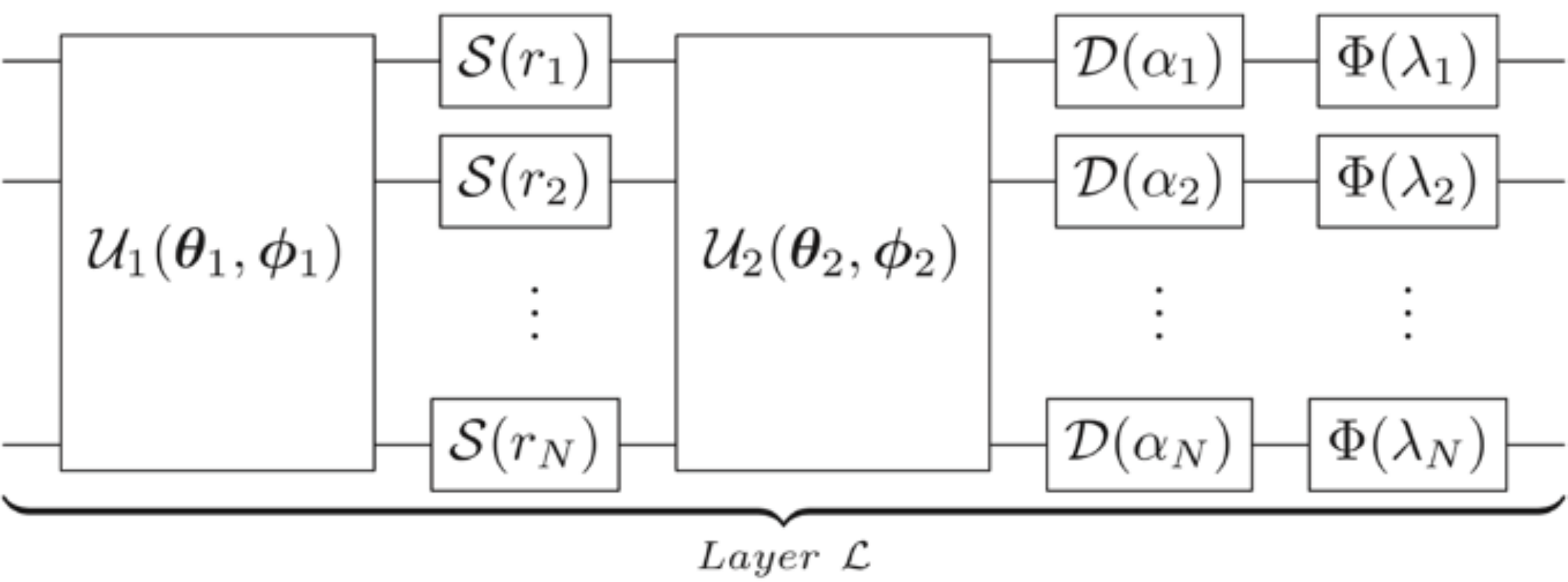}
    \caption{CV quantum neural network architecture \cite{cv_19}}
    \label{fig:qnn}
\end{figure}

The interferometer $U_k$ on an $m-$qumode system is composed of $m-1$ beamsplitters and $m$ rotation gates as shown in the figure. 

\begin{figure}[H]
    \centering
    \includegraphics[scale=0.4]{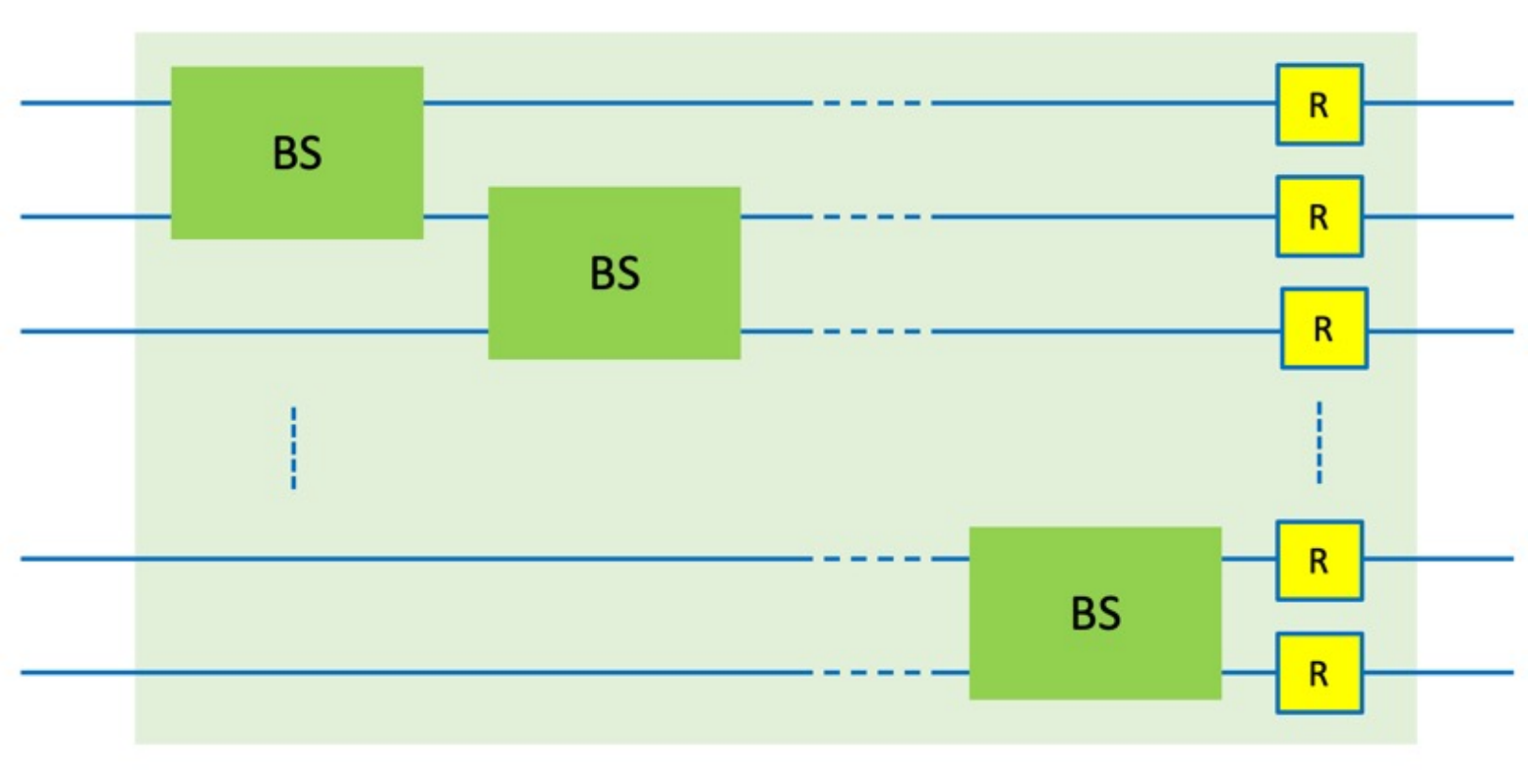}
    \caption{Make-up of an interferometer}
    \label{fig:inter}
\end{figure}

The action of a phaseless interferometer $\mathit{U}_k$ on the quantum state $\ket{x} = \otimes_{k=1}^m \ket{x_k}$ has an effect of an orthogonal matrix acting on $\ket{x}$ \cite{cv_19}. Orthogonal matrices are just unitary matrices with real entries, inducing length-preserving rotations. Then the transpose of an orthogonal matrix represents the reverse rotation of the original matrix, thus orthogonal. Then the composition $U_2 \circ S \circ U_1$ can be considered as the composition $O_2 \circ S \circ \left(O_1^T \right)^T$, where $O_2$ and $O_1^T$ are orthogonal.

Let $W$ be the linear transformation matrix we want to implement with a quantum circuit. Any matrix $W$ can be factorized using singular value decomposition (SVD) as $W = U \Sigma V^{\ast}$, where $U$ and $V$ are orthogonal and $\Sigma$ is diagonal \cite{SVD}. The parameterized squeezer $S(r_k)$ acts on the quantum state $\ket{x_k}$ of each $k^{th}$ qumode as $S(r_k)\ket{x_k} = \sqrt{e^{-r_k}}\ket{e^{-r_k} x_k}$. Collectively they have an effect of a diagonal matrix $\mathit{S}= S(r_1) \otimes S(r_1) \otimes \ldots \otimes S(r_m)$ acting on $\ket{x} = \otimes_{i=1}^m \ket{x_k}$. The composition $\mathit{U}_2 \circ \mathit{S} \circ \mathit{U}_1$ implements a quantum version of the linear transformation matrix $W$ \cite{cv_19}. 

The bias addition is realized with displacement gates $D$. 
The displacement gate has an effect 
\[
D(\alpha_k )\ket{\psi_k}=\ket{\psi_k +\sqrt{2}\alpha_k}
\]
for each $k^{th}$ qumode. Then $D(\alpha)\ket{\psi}=\ket{\psi +\sqrt{2}\alpha}$ collectively for $\alpha^T = [\alpha_1, \alpha_2, \ldots, \alpha_m]$. For some desired bias $b$, let $\alpha = \frac{b}{\sqrt{2}}$, then the collection of displacement gates implements the bias addition. The composition $D \circ U_2 \circ S \circ U_1$ acting on the quantum state $\ket{x}$ gives us the affine transformation 
\[
D \circ U_2 \circ S \circ U_1 \ket{x}=\ket{O_2 \Sigma O_1 x + b}=\ket{Wx+b}.
\]
The non-linear activation function $\phi(\cdot)$ is realized with Kerr gates. The Kerr gate, parameterized by the parameter $\kappa$, is a non-linear transformation gate. Let $n$ be the cutoff dimension and $m$ the number of qumodes. For the quantum state $\ket{\psi}$ of one qumode, which is a superposition of $n$ Fock basis states, the Kerr gate with parameter $\kappa$ has an effect

\[
\begin{aligned}
K(\kappa)\ket{\psi} &=
\begin{bmatrix} 
e^{i \kappa 0^2} & 0 &  \ldots &  0\\
0 & e^{i \kappa 1^2} &  \ldots &  0\\
\vdots &  &\ddots &  \vdots\\
0 & 0 &\ldots &  e^{i \kappa (n-1)^2}\\
\end{bmatrix}
\begin{bmatrix} 
\psi_0\\
\psi_1\\
\vdots \\
\psi_{n-1}\\
\end{bmatrix}\\
&=
\begin{bmatrix} 
\psi_0\\
e^{i \kappa 1^2}\psi_1\\
\vdots \\
e^{i \kappa (n-1)^2}\psi_{n-1}\\
\end{bmatrix},
\end{aligned}
\]
which is non-linear.

Together, the circuit $L=\Phi \circ D \circ U_2 \circ S \circ U_1$ gives us a quantum version $L \left( \ket{x} \right)=\ket{\Phi(Wx+b)}$ of a classical neural network $L(x)=\Phi(Wx+b)$.

\subsection{CV Binary Classifier}
The binary classifier outlined in "Continuous-variable quantum neural networks" is a classical and quantum hybrid network \cite{cv_19}. 

The dataset used contains 284,806 genuine and fraudulent credit card transactions with 29 features, out of which only 492 are fraudulent. The dataset is truncated to 10 features as per the paper and 1,968 samples with 1:3 ratio of fraudulent vs. genuine. 

The proposed classical-quantum hybrid model has a classical neural network taking input vectors of size 10 and outputting vectors of size 14, quantum encoding circuit, and a 2-qumode quantum neural network which outputs vectors of size 2. We can regard the output vectors as one-hot encoding of binary classification of fraudulent vs. genuine. The architecture of the hybrid network is
\begin{figure}[H]
    \centering
    \includegraphics[scale=0.4]{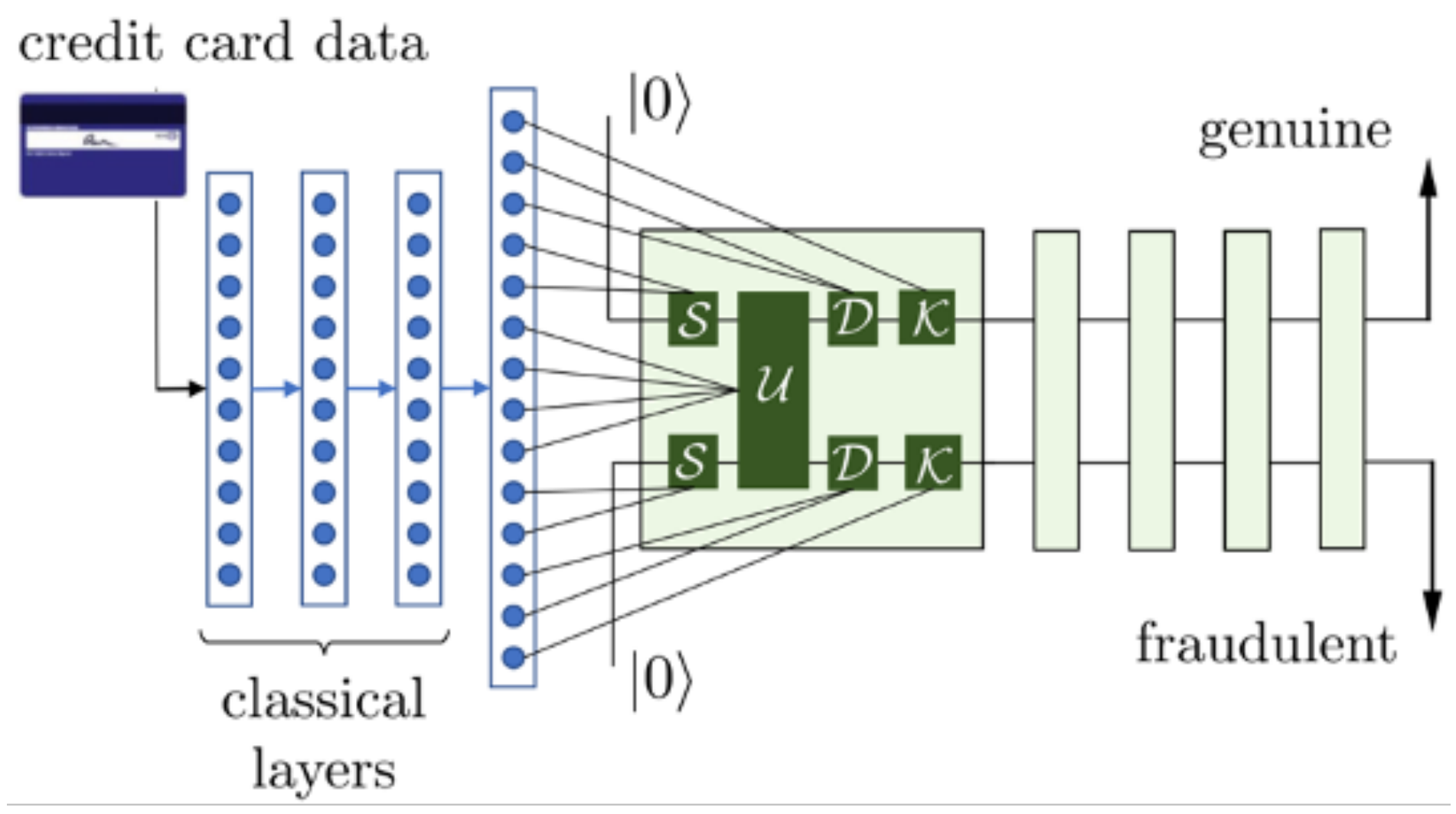}
    \caption{Binary hybrid classifier circuit \cite{cv_19}}
    \label{fig:binary}
\end{figure}

The data flow of the circuit is 
\begin{itemize}
    \item Classical network: 2 hidden layers with 10 neurons, each using Exponential Linear Units (ELU) as activation function. Output layer with 14 neurons.
    \item Data encoding: Output vector from the classical network is converted to a quantum state by the circuit - squeezers, interferometer, displacement gates, and Kerr gates
    \item Quantum network: 4 layers of QNN. Each layer is composed of interferometer 1, squeezers, interferometer 2, displacment gates, and Kerr gates. 
    \item Measurement: The expectation value of the Pauli$-X$ gate $\bra{\phi_k}X \ket{\phi_k}$ is evaluated for each qumode state$ \ket{\phi_k}$ for the $k^{th}$ qumode.
\end{itemize}

The experiment yields $97\%$ training accuracy.

\href{https://github.com/sophchoe/Binary_Classification_Pennylane_Keras}{Code: Keras-Pennylane implementation}

%% file: exp.tex
\section{CV MNIST CLASSIFIERS}

The multi-classifier models presented in this section are inspired by the CV binary classifier architecture described in the section above \cite{cv_19}. They are run on Xanadu's $X8$ photonic quantum simulator, made up of $8-$qumodes. There are two models classifying 8 classes: digits $0, 1, \ldots, 7$ and five models classifying 10 classes: digits $0, 1, \ldots, 9$. The measurement outputs from these models are interpreted as one-hot encoded predictions of image labels.

The classifiers on 8 classes are:

\begin{tabular}[h]{c| ccc}
\hline
num of  & cutoff & measurement & output \\ 
qumodes  &  dim & method & size \\ \hline
3 &  2  & probability & $2^3=8$ \\ 
8 &  2  & expectation value &$1 \times 8 = 8$ \\ \hline
\end{tabular}

A classical neural network is used as a pre-processing step to reduce the original image matrices of size $28 \times 28 = 784$ to vectors of lower length that the data encoding circuit can accommodate. For the data encoding circuit, squeezers, an interferometer, displacement gates, and Kerr gates are used. For the quantum neural network, the quantum circuit implementing $L(\ket{x})=\ket{\phi(Wx+b)}$ as in the binary classifier is used. 

The 10-class classifiers are built on $2, 3, \ldots, 6-$qumodes. The label $k \in \{0, 1, \ldots, 9\}$ of an image matrix, when converted into a one-hot encoded vector, becomes a vector of length $10$ with all zeros but the $k^{th}$ entry as $1$. The cutoff dimensions are selected so that the output vectors exceed $10$ in length.  For each classifier, a different number of zeros are padded to the one-hot encoded labels to match the output size of the circuit. The cutoff dimension used for each classifier and the size of output vectors are shown in the table.

\begin{tabular}[h]{c| ccc}
\hline
num of   & cutoff  &output & number of   \\ 
qumodes   &  dim & size & padding 0's \\ \hline
2 & 4 & $4^2=16$  & 6\\
3 & 3 & $3^3=27$ & 17\\
4 & 2 & $2^4=16$ & 6\\
5 & 2 & $2^5=32$ & 22 \\
6 & 2 & $2^6=64$ & 54\\
\hline
\end{tabular}

The architecture in Figure 6 depicts the data flow of image matrix $\rightarrow$ classical layers $\rightarrow$ reduced output vectors $\rightarrow$ data encoding $\rightarrow$ QNN $\rightarrow$ measurement $\rightarrow$ output vectors as one-hot encoded labels.

\begin{figure}[H]
    \centering
    \includegraphics[scale=0.5]{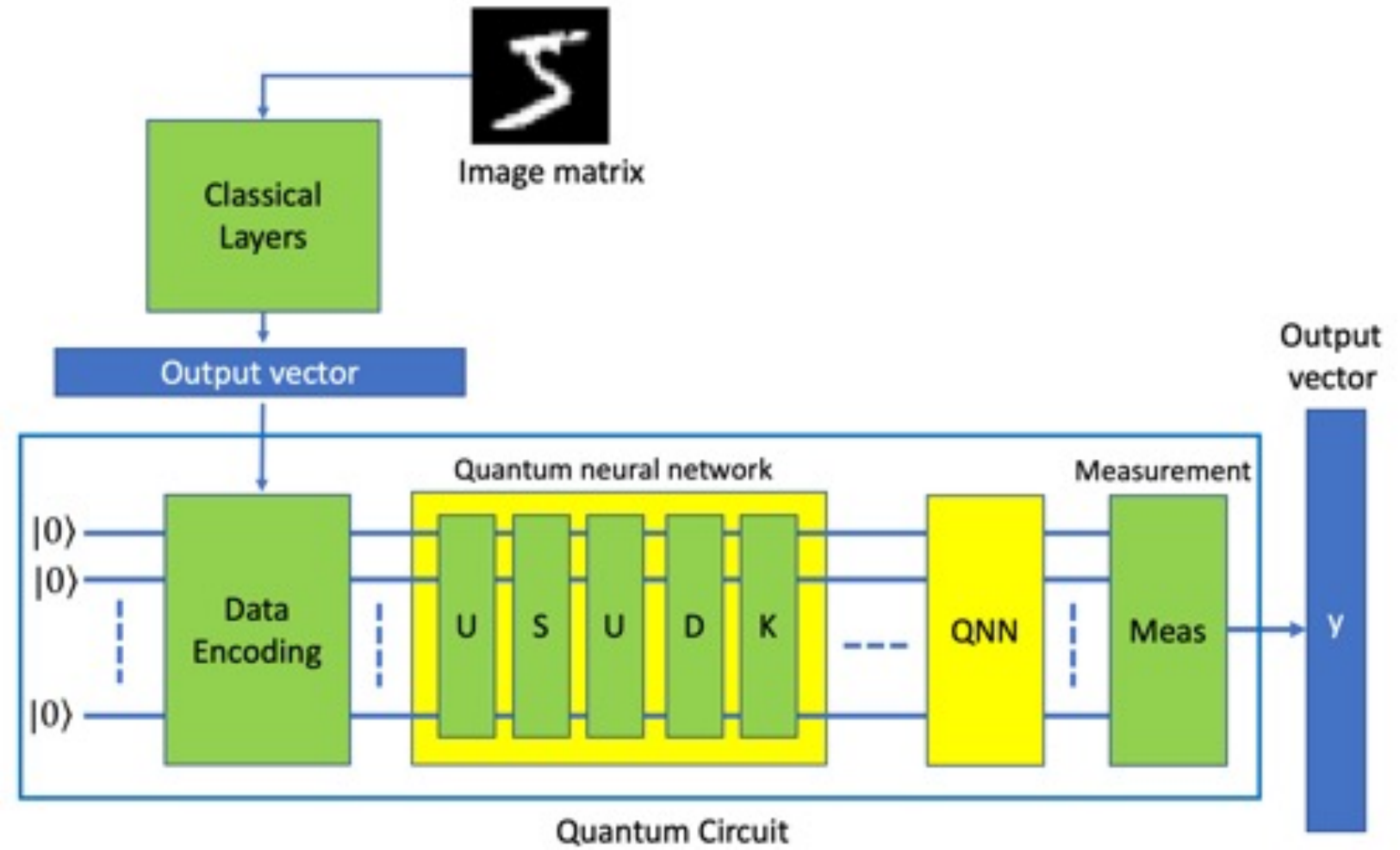}
    \caption{MNIST multi-classifier architecture}
    \label{fig:arch}
\end{figure}

The boxes $U, S, D,$ and $K$ represent interferometers, a set of $m$ squeezers, a set of $m$ displacement gates, and a set of $m$ Kerr gates respectively, where $m=$the number of qumodes. 

\subsection{Classical layers}
Classical feed forward neural networks are used for pre-processing of data images to reduce the size of $28 \times 28 = 784$ to smaller size vectors fitting the number of parameters available for data encoding. The image matrices are flattened to vectors of size $28 \times 28 = 784$ and then reduced to vectors of smaller sizes through Keras dense layer operations with activation function "ELU". The output vectors are then encoded as quantum states by the data encoding quantum circuit. 

\subsection{Data encoding}
The output vectors from the classical neural network are in classical states. The quantum data encoding circuit converts classical states into quantum states. The quantum gates used for data encoding are squeezers, interferometer, displacement gates, and Kerr gates. The entries of a classical vector are used as the parameters of these parameterized quantum gates. 

An interferometer is composed of beamsplitters on pairs of adjacent qumodes and rotation matrices. Squeezers $S(z)$ and displacement gates $D(\alpha)$ can be considered either one parameter gates or two parameter gates when we view the parameters $z, \alpha \in \mathbb{C}$ in Euler formula $z = a + bi = r \left(\cos{\phi} + i \sin{\phi} \right) = re^{i\phi}.$ For the purpose of data encoding, we use them as two-parameter gates. 

The number of parameters that these gates can accommodate for $m-$qumode circuits are $8m - 2$, where $m$ is the number of qumodes.

\begin{tabular}[H]{ccccc}
\hline
squeezers  & BS & rotation & displacement & Kerr  \\  \hline
$2m$ & $2(m-1)$ & $m$ & $2m$ & $m$ \\ \hline
\end{tabular}

Based on these values, the size of the classical network output vectors was determined.

\subsection{Quantum circuit}
The QNN circuit implements a quantum version of the classical neural network
$L(x) = \phi(Wx+b)$ as 
\[
L(\ket{x})=\ket{\phi(Wx+b)}=\phi \circ D \circ U_2 \circ S \circ U_i \ket{x}
\]
where $\ket{x}$ is a quantum data encoded state of the original data vector $x$, $U_k$ interferometers, $S$ squeezers, $D$ displacement gates, and $\phi(\cdot)$ Kerr gates respectively. The number of parameters per a collection of gates on $m-$qumodes are

\begin{tabular}[H]{cccc |c}
\hline
$U_k$  & $S$ & $D$ & $\phi(\cdot)$ & total \\  \hline
$2(m-1)+m$ & $m$ & $m$ & $m$ & $2 \cdot 2(m-1)+5m$\\
 &  & &  & $= 9m - 4$\\ \hline
\end{tabular}

On the 8-qumode classifier, 2 layers of QNN are applied. On the rest of classifiers, 4 layers are applied. The number of parameters for the varying number of qumodes are shown in the table.

\begin{tabular}[H]{c|cc}
\hline
num of qumodes & per layer & total \\  \hline
2 & 14 & 56 \\ 
3 & 23 & 92 \\ 
4 & 32 & 128 \\ 
5 & 41 & 164 \\ 
6 & 50 & 200\\ 
8 & 68 & 136 \\ \hline
\end{tabular}

\subsection{Measurement}
Pennylane offers three different measurement methods for CV-based computations: expectation value, variance, and probabilities. The expectation value and variance methods yield a single-valued result for each qumode. The probability method yields vectors of size $n^m$, where $n=$cutoff dimension and $m=$the number of qumodes.

 For the eight-qumode model, the expectation value method was used. The result is a vector of length $8: [\bra{\psi_0}X\ket{\psi_0}, \bra{\psi_1}X\ket{\psi_1}, \ldots, \bra{\psi_7}X\ket{\psi_7}]$ where $\bra{\psi_k}X\ket{\psi_k}$ represents the expectation value measurement result of the $k^{th}$ qumode. It is then translated as a one-hot encoded label vector of the corresponding image. 
 
For the rest of the models, the probability method was used. 

\subsection{Parameter update}
With the Pennylane Tensorflow plug-in functionality, the quantum circuit is converted into a Keras layer and added to the classical layers. Then Keras' built-in loss functions and optimizers can be used for parameter update. For most of the models, Categorical Crossentropy is used for loss function and Stochastic Gradient Descent is used for optimizer. For the 8-qumode classifier, the Mean Squared Error loss function performed better. The updated parameters are then used as the parameters of the quantum gates for the subsequent iteration of training.

\subsection{Experimetal results}

The $4-$qumode classifier yielded the best result of $100\%$ training accuracy on $600$ data samples. Accuracy comparison with the qubit-based binary classifiers using Tensorflow Quantum and Qiskit is listed in the table below. 

\scalebox{0.72}{
\begin{tabular}{ |c|c|c|c|}
\hline
    & Tensorflow Q & Qiskit - PyTorch & Pennylane - Keras \\ \hline
Number of classes   & 2   & 2 & 10     \\ \hline
Number of samples   & 10,338 & 100 & 600     \\ \hline
Training accuracy   & $89.92\%$ & $100\%$ & $100\%$   \\ \hline
\end{tabular}
\label{Tab:MNIST}}

All the classifiers tested achieve above $95\%$ training accuracy. 
For the 8-qumode classifier, 300 samples and 2 layers of QNN were used with 50 epochs. For the rest of the classifiers, 600 samples and 4 layers of QNN were used. With the 4-qumode classifier, training accuracy of $100\%$ is achieved in 70 epochs. 

The loss and accuracy for the models with varying number of qumodes are listed below.

\begin{tabular}[h!]{c| ccc}
\hline
num of & learning & training & training \\ 
qumodes & rate &  loss &  accuracy \\ \hline
2 & 0.02 & 0.4966 & $97.14\%$ \\
$3^{\ast}$ & 0.05 & 0.0563 & $99.90\%$   \\ 
3 & 0.05 & 0.4567 & $98.04\%$\\
4 & 0.03 & 0.1199 & $100.00\%$\\
5 & 0.02 & 0.2057 & $98.40\%$ \\ 
6 & 0.02 & 0.2055 & $98.22\%$\\
8 & 0.1 & 0.0199 & $99.77\%$ \\\hline
\end{tabular}

The second line $3^{\ast}$ indicates an $8-$class classifier. 

All of these classifiers followed the typical loss and accuracy graphs depicted in Figure 7. The validation accuracy is not as high as the training accuracy, indicating "over-fitting". Further experiments with different hyper-parameters or the use of regularizer are called for. 

\href{https://github.com/sophchoe/Continous-Variable-Quantum-MNIST-Classifiers}{Code: Keras-Pennylane implementation}

\begin{figure}[H]
    \centering
    \includegraphics[scale=0.5]{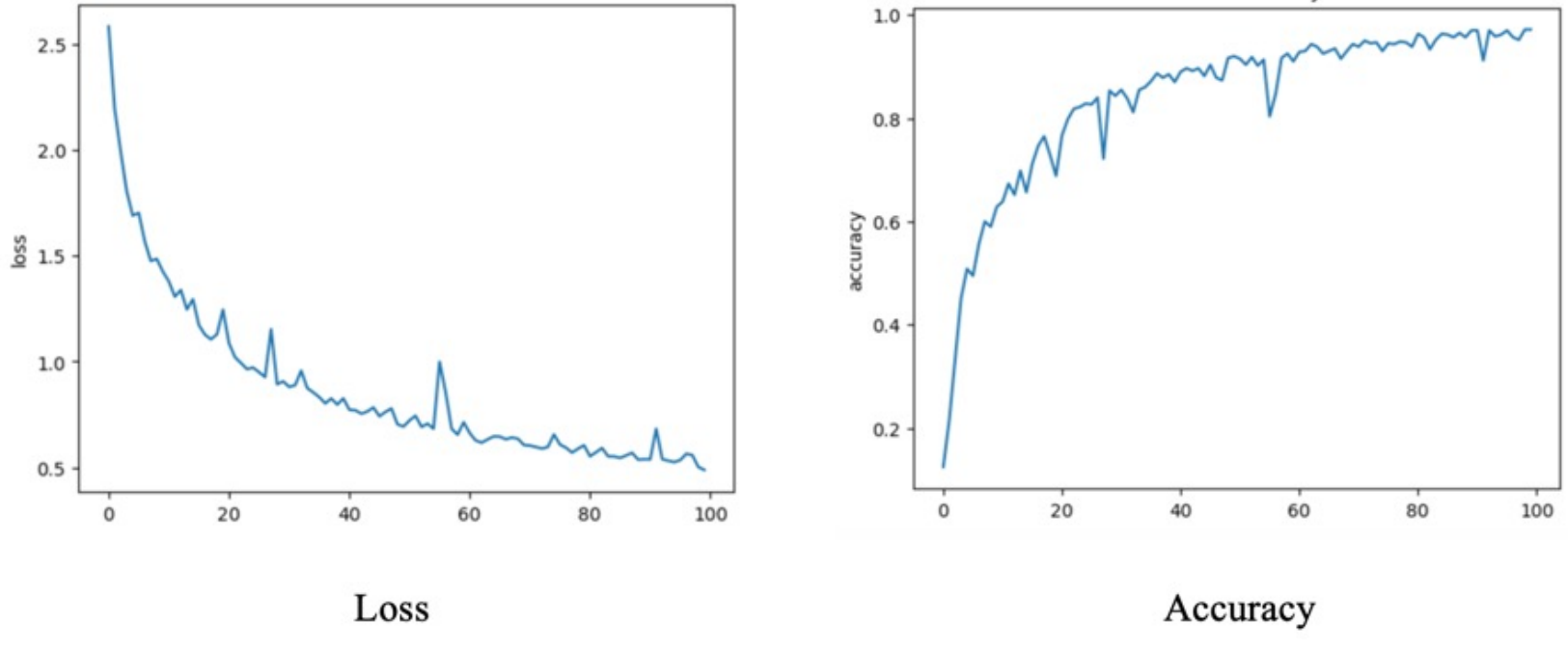}
    \caption{MNIST multi classifier experimental results}
    \label{fig:MNIST_exp}
\end{figure}

%% file: conc.tex
\section{CONCLUSION}
In this pape, classical and CV quantum hybrid classifiers using different number of qumodes and cutoff dimensions were examined. The quantum gates available in the CV model allow a natural implementation of a quantum neural network layer $L(\ket{x})=\ket{\phi(Wx+b)}$. The flexibility of different measurement methods to output vectors of different lengths allows the networks to produce results that are interpreted as one-hot encoded labels of MNIST image dataset. 

There is a limitation in encoding classical data into quantum states on near-term devices due to the number of qumodes, which is currently eight on Xanadu's X8 photonic QPU. Classical networks were used to reduce the number of entries in the image matrices for quantum data encoding. Although the role of the classical network is for pre-processing, the majority of parameters that are learned through the training process is on the classical network side. One way of encoding all the entries of an image matrix would be iterating through smaller sections of the image, which naturally segues to convolutional operations. A combination of quantum convolutional network layers and quantum neural network layers is a way of implementing a purely quantum network.

In implementing machine learning algorithms in quantum, the CV model offers many advantages over the qubit-based model. The quantum computational state space of the CV model is infinite while it is finite in the qubit-based model. The ability to define the dimension of the CV quantum computational space in approximating the original infinite computational space avails users of added freedom and flexibility in experimenting quantum algorithms. Photonic quantum computers, which are a version of physical implementation of the CV model can be easily incorportated into the current computing infrastructure because of their operability at room temperature.